\documentclass[onecolumn,aps,prd,preprintnumbers,showpacs,superscriptaddress,nofootinbib,amsmath,amssymb,floats,floatfix,showkeys,notitlepage,longbibliography]{revtex4-1}

\usepackage{comment}
\usepackage{lipsum}
\usepackage{graphicx}
\usepackage{subfigure}
\usepackage{palatino}
\usepackage{sans}
\usepackage{hyperref}
\hypersetup{colorlinks=true,linkcolor=blue,urlcolor=blue,citecolor=blue}
\usepackage[toc,page]{appendix}
\usepackage[normalem]{ulem}
\usepackage{adjustbox}
\usepackage{latexsym}
\usepackage{amsmath}
\usepackage{amssymb}
\usepackage{amsfonts}
\usepackage{dcolumn}
\usepackage{bm}
\usepackage{tikz}
\usepackage{bigints}
\usepackage{array,tabularx,multirow,booktabs}
\usepackage[tracking=true]{microtype}
\usepackage{soul} 
\SetTracking{}{500}
\SetTracking{encoding={*}, shape=sc}{40}
\UseRawInputEncoding 
\allowdisplaybreaks

\begin{document} \sloppy
\title{Constraints via EHT for black hole solutions with dark matter under the generalized uncertainty principle minimal length scale effect}

\author{Ali \"Ovg\"un}
\email{ali.ovgun@emu.edu.tr}
\affiliation{Physics Department, Eastern Mediterranean
University, 99628, Famagusta, North Cyprus
via Mersin 10, Turkey.}

\author{Lemuel John F. Sese}
\email{ljfsese@mapua.edu.ph}
\affiliation{Physics Department, Map\'ua University, 658 Muralla St., Intramuros, Manila 1002, Philippines.}

\author{Reggie C. Pantig}
\email{rcpantig@mapua.edu.ph}
\affiliation{Physics Department, Map\'ua University, 658 Muralla St., Intramuros, Manila 1002, Philippines.}

\begin{abstract}
Four spherically symmetric but non-asymptotically flat black hole solutions surrounded with spherical dark matter distribution perceived under the minimal length scale effect is derived via the generalized uncertainty principle. Here, the effect of this quantum correction, described by the parameter $\gamma$, is considered on a toy model galaxy with dark matter and the three well-known dark matter distributions: the cold dark matter, scalar field dark matter, and the universal rotation curve. The aim is to find constraints to $\gamma$ by applying these solutions to the known supermassive black holes: Sagittarius A (Sgr. A*) and Messier 87* (M87*), in conjunction with the available Event Horizon telescope. The effect of $\gamma$ is then examined on the event horizon, photonsphere, and shadow radii, where unique deviations from the Schwarzschild case are observed. As for the shadow radii, bounds are obtained for the values of $\gamma$ on each black hole solution at $3\sigma$ confidence level. The results revealed that under minimal length scale effect, black holes can give positive (larger shadow) and negative values (smaller shadow) of $\gamma$, which are supported indirectly by laboratory experiments and astrophysical or cosmological observations, respectively.
\end{abstract}


\maketitle


\section{Introduction} \label{intr}

Black holes (BHs) are one of the most remarkable predictions of the theory of General Relativity (GR), offering a unique opportunity to explore theories of gravity through various strong-field phenomena, such as the formation of black hole shadows \cite{Bardeen:1973tla}. Recently, the Event Horizon Telescope collaboration made a groundbreaking achievement by publishing the first-ever image of a black hole (Event Horizon Telescope Collaboration 2019) \cite{EventHorizonTelescope:2019dse,EventHorizonTelescope:2022xnr}. This image exhibits an intriguing effect of a constant spacetime structure being illuminated by a time-varying emission region, leading to a need for a deeper understanding and explanation of the observed phenomena. The visual appearance of black holes has been a subject of study since Cunningham $\&$ Bardeen \cite{Cunningham} and Bardeen \cite{1974IAUS...64..132B} first investigated the case of a star orbiting a black hole, along with other related scenarios. Their early work revealed the fundamental structure of a thin disk with an inner region gravitationally lensed. Luminet (1979) \cite{Luminet:1979nyg} subsequently presented the first computer-calculated visualization of a black hole surrounded by a luminous accretion disk, although it was hand-drawn at that time. The term "shadow" was introduced independently by Falcke et al. \cite{Falcke:1999pj} and de Vries \cite{deVries:1999tiy} a few weeks later. Subsequently, the Event Horizon Telescope (EHT) has adopted this term to describe the observable phenomenon in black hole imaging.

The true nature of dark matter remains one of the greatest mysteries in astrophysics and cosmology. The $\Lambda$CDM model, which successfully explains the dynamics of the large-scale Universe, suggests that dark matter constitutes approximately $27\%$ of the Universe's content and about $85\%$ of its total mass \cite{WMAP:2010sfg}. However, the model faces significant challenges at the galactic scale, including the cusp-core problem, the missing satellite problem, and the too-big-to-fail problem \cite{Moore:1994yx, Klypin:1999uc, Boylan-Kolchin:2011qkt}.

One elusive aspect is the Earth-based detection of dark matter particles known as WIMPs (Weakly Interactive Massive Particles), which are associated with the $\Lambda$CDM model. Some studies initially reported positive results \cite{ Bernabei:2018jrt}, but subsequent experiments from other testing laboratories provided null results, debunking these findings \cite{CRESST:2015txj, PICO:2017tgi}. Other proposed alternatives involve studying the Earth's crust for years of data that might reveal imprints of dark matter \cite{Baum:2018tfw}. As of the latest update, even the most sensitive dark matter detector reported no detection of dark matter particles \cite{LZ:2022lsv}.

Interestingly, black holes exhibit traces not only of their intrinsic spacetime geometry but also of the exotic matter surrounding them, providing valuable insights into different matter-energy distributions in their immediate vicinity \cite{2013A&A...554A..36L,Kumar:2017tdw,Atamurotov:2021cgh}. Moreover, the BH shadow serves as a powerful tool to constrain the otherwise elusive surrounding matter distribution \cite{Ma:2020dhv,Saurabh:2020zqg,Das:2020yxw,Atamurotov:2021hck,Nampalliwar:2021tyz,Mustafa:2023kqt,Rakhimova:2023rie}, encompassing not only luminous matter but also the signatures of dark matter.

In recent research, the impact of the Dehnen profile on specific black holes situated in a dwarf galaxy was investigated in Ref. \cite{Pantig:2022whj}. Then, the effect of the fuzzy dark matter (or wave dark matter) halo on a supermassive black hole was explored in Ref. \cite{Pantig:2022sjb}.  Additionally, Cardoso et al. conducted a study on the evolution of a fuzzy dark matter soliton during its accretion by a central supermassive black hole. They identified various stages of accretion and the corresponding timescales involved in the process \cite{Cardoso:2022nzc}. Konoplya \cite{Konoplya:2019sns,Konoplya:2022hbl} examined a toy model of dark matter and investigated its influence on the black hole shadow's effective mass.  Moreover, other dark matter profiles, including CDM, SFDM, URC, and superfluid dark matter, were analyzed in relation to their influence on black hole geometry in various references (Hou et al. \cite{Hou:2018bar,Hou:2018avu}). Furthermore, more intricate models of dark matter profiles have been explored (Xu et al. \cite{Xu:2020jpv,Xu:2018wow,Xu:2018mkl}), along with profiles featuring dark matter spikes (Nampalliwar et al. \cite{Nampalliwar:2021tyz}, Xu et al. \cite{Xu:2021dkv}). Anjum et al. have studied shadow of the Kerr black holes (BHs) surrounded by perfect fluid dark matter \cite{Anjum:2023axh}. In addition, various effects of perfect fluid dark matter were explored in Refs. \cite{Rayimbaev:2021kjs,Shaymatov:2020wtj,Hendi:2020zyw, Rizwan:2018rgs} such as particle epicyclic motion, instability and phase transition, testing the weak cosmic censorship, and using precession frequency in distinguishing naked singularity to Kerr-like black holes.

In this study, we will use an alternative perspective on explaining dark matter that some dark matter phenomena could be understood partially through quantum gravitational effects. Developing a complete Quantum Gravity (QG) theory remains an unsolved challenge, with various theoretical frameworks proposed. However, direct testing of these theories is limited by current experimental constraints. To address this, phenomenological methods have gained prominence \cite{Addazi:2021xuf,Das:2021skl,Jizba:2023ygi,Chen:2022kzv}, where the influence of QG concepts on lower energy scales is explored, offering insights amenable to current observations.

One key QG notion is the introduction of a fundamental minimal length scale, which challenges the traditional Heisenberg Uncertainty Principle. This gives rise to the Generalized Uncertainty Principle (GUP) \cite{Maggiore:1993zu,Kempf:1994su,Scardigli:1999jh,Lambiase:2023hng,Lambiase:2022xde,Okcu:2021oke,Jusufi:2020wmp}, encompassing models inspired by quantum gravity candidates like string theory and loop quantum gravity. These models introduce an effective minimal observable length, leading to implications such as minimal position uncertainty or spacetime non-commutativity. This approach paves the way for understanding DM through quantum gravitational insights. The commutator that characterizes one of the most prevalent GUP models can be mathematically expressed as \cite{Kempf:1994su}

\begin{equation}
\left[x_{i}, p_{j}\right]=i \hbar\left[\delta_{i j}+\beta\left(p^{2} \delta_{i j}+2 p_{i} p_{j}\right)\right].
\end{equation}

Here, \(x_{i}\) and \(p_{j}\) denote the position and momentum operators, respectively. The parameter \(\beta\) is defined as \(\beta \equiv \beta_{0} /\left(M_{P} c\right)^{2}\), where \(\beta_{0}\) is a dimensionless constant and \(M_{P}\) represents the Planck mass defined as \(\sqrt{\hbar c / G}\). This formulation leads to a revision of the uncertainty relationship between position and momentum. This revision becomes apparent when considering the Schrödinger-Robertson relation, which ultimately results in a minimal uncertainty in position. To provide an instance, in the context of the one-dimensional scenario, the modified uncertainty relation is given by  \cite{Kempf:1994su} \begin{equation}
\Delta x \Delta p \gtrsim \hbar\left[1+3 \beta \Delta p^{2}\right].
\end{equation}

This study explores the potential connections between minimal-length phenomenology and the emergence of phenomena resembling dark matter (DM) on galactic scales. Specifically, we will leverage the findings presented in the work by Bosso et al. \cite{Bosso:2022xnm}, where the effects arising from the Generalized Uncertainty Principle (GUP) were identified as contributors to the observed flatness of rotational curves. We will employ these results to construct a black hole solution incorporating these insights. Our study aims to examine the shadow patterns of black hole solutions incorporating dark matter, considering the influence of the generalized uncertainty principle and the minimal length scale effect. We aim to uncover significant insights and offer guidance for the experimental identification of this specific spacetime arrangement by future space technologies, far more sophisticated than the EHT. Such an inquiry will significantly augment our comprehension of black hole characteristics and facilitate the interpretation of black hole shadow observations within the framework of astrophysical scenarios.

We organize the paper as follows: In Sect. \ref{sec2}, we derive the metric of the black holes in dark matter halo with additional effect from the minimal length scale from GUP. In Sect. \ref{sec3}, we theoretically study the photonsphere and shad behavior of each black hole solution. We also find constraints to the GUP parameter $\gamma$ using the EHT data for Sgr. A* and M87*. Finally, in Sect. \ref{sec4}, we summarize the results and propose a research direction. Throughout the paper, the metric signature is $(-,+,+,+)$ and geometrized units are used imposing $G = c = 1$.

\section{Derivation of the black hole metric} \label{sec2}
In this section, we aim to derive the black hole metric using the algorithm developed in Ref. \cite{Xu:2018wow}. Here, we will extend the algorithm by adding the GUP effect as a minimal length scale and study its consequences on certain black hole properties.

We begin with the exponential density profile \cite{Freeman:1970mx,Bosso:2022xnm} 
given as
\begin{equation} \label{e1}
    \rho(r) = \rho_0 e^{-r/r_d},
\end{equation}
where $\rho_0 = 2$x$10^{-19} \text{kg/m}^3$ as the central density, then $r_d = 19000$ ly as the galaxy scale parameter \cite{Freeman:1970mx,Bosso:2022xnm}. These values translates to $\rho_0 = 2.955$x$10^{9} M_\odot\text{/kpc}^3$ and $r_d = 5.825$ kpc. While Eq. \eqref{e1} can be viewed as a toy model, the resulting behavior of the velocity profile exhibits similarity to the velocity profile derived in Ref. \cite{Freeman:1970mx}. In that study, the radial distribution of the surface brightness $I(R)$ of the disks of some sample galaxies follows an exponential law: $I(R) = I_0e^{-aR}$. It implied that the surface-density distribution, which is the distribution of matter, should also follow an exponential law \cite{Freeman:1970mx}.

We could then find the mass $\mathcal{M}$ of the distribution through the formula
\begin{equation}
    \mathcal{M}(r) = 4 \pi \int_{0}^{r} \rho\left(r^{\prime}\right) r^{\prime 2} d r^{\prime},
\end{equation}
resulting to
\begin{equation}
    \mathcal{M}(r) = -4 \pi r_d \left[\left(r^{2}+2 r r_d +2 r_d^{2}\right) {\mathrm e}^{-\frac{r}{r_d}}-2 r_d^{2}\right].
\end{equation}
Next, the GUP-modified velocity at large distances takes the form \cite{Bosso:2022xnm}
\begin{equation} \label{vtan}
    v_\text{tg} = \sqrt{\frac{\mathcal{M}}{r}\left(1 - \omega^2\beta(r) \right)},
\end{equation}
where we have written $\omega = \frac{m/n}{m_{Pl}}$. Note that $m/n$ is the order of the proton's mass for any value of $m_p$ and $n$, $m_{Pl}$ is the Planck's mass. Thus, one can safely assume that the variable $\omega$ is a small number. Now, $\beta(r)$ is some parameter that is distance-dependent \cite{Bosso:2022xnm}, where such a claim is supported by several studies in the literature \cite{Ong:2018zqn, Pikovski:2011zk,Scardigli:2016pjs,Bosso:2016ycv,Kumar:2017cka,Das:2021nbq,Nenmeli:2021orl,Jizba:2009qf,Buoninfante:2019fwr,Jizba:2022icu}. In this study, we are interested in its form as \cite{Bosso:2022xnm}
\begin{equation} \label{ebeta}
    \beta(r) = \gamma \frac{r^2}{r_*^2},
\end{equation}
where a constant GUP parameter $\gamma$ is introduced, and $r_* \sim 1$ ly is the scale at which the dark matter effects become significant. With this form of $\beta$ and for brevity, we update Eq. \eqref{vtan} as
\begin{equation} \label{vtan2}
    v_\text{tg} = \sqrt{\frac{\mathcal{M}}{r}\left(1 - \chi r^2 \right)},
\end{equation}
where
\begin{equation} \label{chi}
    \chi = \gamma \frac{\omega^2}{r_*^2}.
\end{equation}
It is clear that $\chi$ should have a geometrized unit of $\text{m}^{-2}$ and be written dimensionless as $\chi m^2$ if expressed in terms of the black hole mass $m$. 

Knowing this tangential velocity, we can consider the line element for the dark matter halo
\begin{equation} \label{hmetric}
    ds^{2}_{\text{halo}} = -f(r) dt^{2} + g(r)^{-1} dr^{2} + r^2 d\theta ^{2} +r^2\sin^2\theta d\phi^{2},
\end{equation}
where we can relate the metric function $f(r)$ to the tangential velocity through \cite{Xu:2018wow}
\begin{equation}
    v_\text{tg}(r)=r \frac{d \ln (\sqrt{f(r)})}{d r}.
\end{equation}
With some integration, we obtain
\begin{equation}
    f(r) = \exp{\left[\frac{8\pi k (2 + r/r_d) e^{-\frac{r}{r_d}} -16\pi k - \chi r^3}{r} \right]},
\end{equation}
where we have written $k = \rho_0 r_d^3$. We remark that for any expression for $\beta(r)$ under Eq. \eqref{e1}, we can find the general formula as
\begin{equation}
    f(r) = \exp\Bigg\{-\frac{8 \pi k}{r_d^{2}} \left[ \int \left(1 - \omega^2 \beta(r) \right) \left[\left(r^{2}+2 r r_d +2 r_d^{2}\right) {\mathrm e}^{-\frac{r}{r_d}}-2 r_d^{2}\right]r^{-2} dr \right] \Bigg\},
\end{equation}
where the possibility of obtaining a closed analytic form/result depends on how one models $\beta(r)$.

Now that we have obtained the metric function of the halo, it is possible to fuse this with the black hole mass $m$ \cite{Xu:2018wow}. The Einstein field equation, with this combination, can be modified as
\begin{equation} \label{e15}
    R^{\mu}_{\nu}=\frac{1}{2}\delta^{\mu}_{\nu}R=\kappa^2((T^{\mu}_{\nu})_{\text{halo}}+(T^{\mu}_{\nu})_{\text{Schw}}),
\end{equation}
which allows us to redefine the metric function as
\begin{equation} \label{e16}
    ds^{2} = -F(r) dt^{2} + G(r)^{-1} dr^{2} + r^2 d\theta ^{2} +r^2\sin^2\theta d\phi^{2},
\end{equation}
where we write
\begin{equation} \label{e17}
    F(r)=f(r) + F_1(r), \quad \quad G(r) = g(r)+F_2(r).
\end{equation}
As a result, Eq. \eqref{e15} gives us
\begin{align} \label{e18}
    (g(r)+F_{2}(r))\left(\frac{1}{r^{2}}+\frac{1}{r}\frac{g^{'}(r)+F^{'}_{2}(r)}{g(r)+F_{2}(r)}\right)&=g(r)\left(\frac{1}{r^{2}}+\frac{1}{r}\frac{g^{'}(r)}{g(r)}\right), \nonumber \\
    (g(r)+F_{2}(r))\left(\frac{1}{r^{2}}+\frac{1}{r}\frac{f^{'}(r)+F^{'}_{1}(r)}{f(r)+F_{1}(r)}\right)&=g(r)\left(\frac{1}{r^{2}}+\frac{1}{r}\frac{f^{'}(r)}{f(r)}\right).
\end{align}
Solving for $F_1(r)$ and $F_2(r)$ yields
\begin{align} \label{e19}
    F(r) &= \exp\left[\int \frac{g(r)}{g(r)-\frac{2m}{r}}\left(\frac{1}{r}+\frac{f^{'}(r)}{f(r)}\right)dr-\frac{1}{r} dr\right], \nonumber\\
    G(r) &=g(r)-\frac{2m}{r},
\end{align}
Note that $f(r)=g(r)=1$ implies the non-existence of the dark matter halo, resulting in the integral of $F(r)$ to become a constant $C_1 = 1-2m/r$. Thus, it merely reduces to the pure Schwarzschild case. The dark matter halo can be found by inspecting Eqs. \eqref{e17}-\eqref{e18}. Then, if we assume that $f(r)=g(r)$ and $F_1(r)=F_2(r)=-2m/r$, it implies that $F(r)=G(r)$, and the metric function can be simply written as 
\begin{equation} \label{dm_metric}
    F(r) = \exp{\left[ \frac{8\pi k (2 + r/r_d) e^{-\frac{r}{r_d}} -16\pi k - \chi r^3}{r}\right]} - \frac{2m}{r}.
\end{equation}
We remark that, in general, for any expression for $\beta(r)$ under Eq. \eqref{e1}, one could find the black hole metric with dark matter halo as
\begin{equation}  \label{emetric}
    F(r) = \exp\Bigg\{-\frac{8 \pi k}{r_d^{2}} \left[ \int \left(1 - \omega^2 \beta(r) \right) \left[\left(r^{2}+2 r r_d +2 r_d^{2}\right) {\mathrm e}^{-\frac{r}{r_d}}-2 r_d^{2}\right]r^{-2} dr \right] \Bigg\}-\frac{2m}{r}.
\end{equation}

Several papers in the literature suggested that for dark matter to manifest its effect on the black hole shadow, the dark matter parameter $k$ should be comparable to the black hole mass $m$ \cite{Konoplya:2019sns}. If $k$ is very small, such an effect can be compensated by a very small core radius $r_d$. There is also some reason to suspect that $\chi$ is a small parameter \cite{Bosso:2022xnm}. Thus, we can apply some series expansion to Eq. \eqref{dm_metric} to make the expression more manageable:
\begin{equation}
    F(r) = 1 - \frac{2m}{r} - \frac{8 \pi k}{r r_d} (r^2 \chi - 1) \left[(2r_d+r)e^{-\frac{r}{r_d}} - d \right] - \chi r^2,
\end{equation}
where we could see a coupling between the dark matter and the GUP parameter, in addition to mimicking the effects of the cosmological constant $\Lambda$. This result agrees with the claim that dark matter can be perceived as a minimal length scale effect from the GUP, in which its dominance manifests at cosmological distances \cite{Bosso:2022xnm}.

We can apply the above results, albeit theoretically, to Sgr. A* and M87* using the parameters mentioned herein to get a glimpse of how the black hole properties would change under the combined effects of dark matter distribution and the GUP parameter. To become more realistic, we will add three more density profiles in this study, where empirical data is readily available for $k$ and $r_d$. These models are the cold dark matter (CDM), scalar field dark matter (SFDM), and the universal rotation curve dark matter (URC), where the density profiles are given by
\begin{equation}
    \rho^\text{CDM} = \frac{\rho_0}{\frac{r}{r_d}\left(1+\frac{r}{r_d}\right)^2}, \quad \rho^\text{SFDM} = \frac{\rho_0 r_d}{\pi r}\sin\left(\frac{\pi r}{r_d}\right), \quad \rho^\text{URC} = \frac{\rho_0 r_d^3}{(r+r_d)(r^2+r_d^2)},
\end{equation}
respectively. 
The CDM model uses the Navarro-Frenk-White (NFW) profile, which is a widely used model for the density distribution of dark matter halos \cite{Navarro:1995iw}. Its most important prediction (confirmed by observation) is the existence of large number of subhalos, which are smaller halos that have merged with the main halo over time \cite{Moore:1999nt,Springel:2008cc}. The SFDM model is a relatively new dark matter model, but it has quickly gained popularity due to its ability to address some of the shortcomings of the standard cold dark matter (CDM) model. For example, the SFDM model can naturally produce flat central density profiles in galaxy halos, consistent with observations. Ref. \cite{Suarez:2013iw} gave a comprehensive overview of the SFDM model, including its theoretical foundations, cosmological implications, and observational constraints. Meanwhile, the URC model is now widely accepted as strong evidence for the existence of dark matter, and it is used by astronomers to study the distribution of dark matter in galaxies \cite{Persic:1995ru}. The URC is a plot of the circular velocity of a galaxy as a function of radius, and it shows that the circular velocity of a galaxy does not drop off as quickly as expected from the visible matter alone. This suggests that there must be an additional, invisible component of mass, known as dark matter, that is contributing to the gravity of the galaxy.

If we add the GUP modification at large distances, we obtain the corresponding black hole solutions while applying the approximations $k << m$, and $\chi \sim 0$:
\begin{equation}
    F(r)^\text{CDM} = 1 - \frac{2m}{r} - \frac{8 \pi k}{r}(1 + \chi r_d^2 - \chi r^2)\ln{\left(1-\frac{r}{r_d}\right)} + (r_d^2 - r^2) \chi,
\end{equation}
\begin{equation}
    F(r)^\text{SFDM} = 1 - \frac{2m}{r} + \frac{8k}{\pi^2 r}\sin\left(\frac{\pi r}{r_d} \right) (\chi ^2 r -1) - \chi r^2,
\end{equation}    
\begin{align}
    F(r)^\text{URC} = 1 - \frac{2m}{r} + \frac{2 \pi k}{3 r_d} (\chi r^2 - 1) \Bigg\{\ln{\left[\frac{(r + r_d)^2}{r^2 + r_d^2}\right]} - 2\arctan(r/r_d) \Bigg\} - \chi r^2.
\end{align}

From now on, we want to write the full metric as
\begin{equation} \label{e21}
    ds^{2} = -A(r) dt^{2} + B(r) dr^{2} + C(r) d\theta ^{2} +D(r) d\phi^{2},
\end{equation}
where $B(r)=A(r)^{-1}$, $C(r)=r^2$, and $D(r)=r^2\sin^2\theta$. The metric coefficients $A(r)$ and $B(r)$ have no dependence on time $(\partial_t g_{\mu\nu} = 0)$ and angular coordinates $\theta$ and $\phi$ $(\partial_\theta g_{\mu\nu} = 0, \partial_\phi g_{\mu\nu} = 0)$, thus, the metric is static and spherically symmetric. It would mean that the gravitational field of the black hole being considered is the same in all directions. Thus, without loss of generality, we can restrict the analysis to $\theta = \pi/2$, leading to a more simplified analysis of a black hole metric in $1+2$ dimensions: $ds^{2} = -A(r) dt^{2} + B(r) dr^{2} + r^2 d\phi^{2}$.

\begin{figure*}
    \centering
    \includegraphics[width=0.48\textwidth]{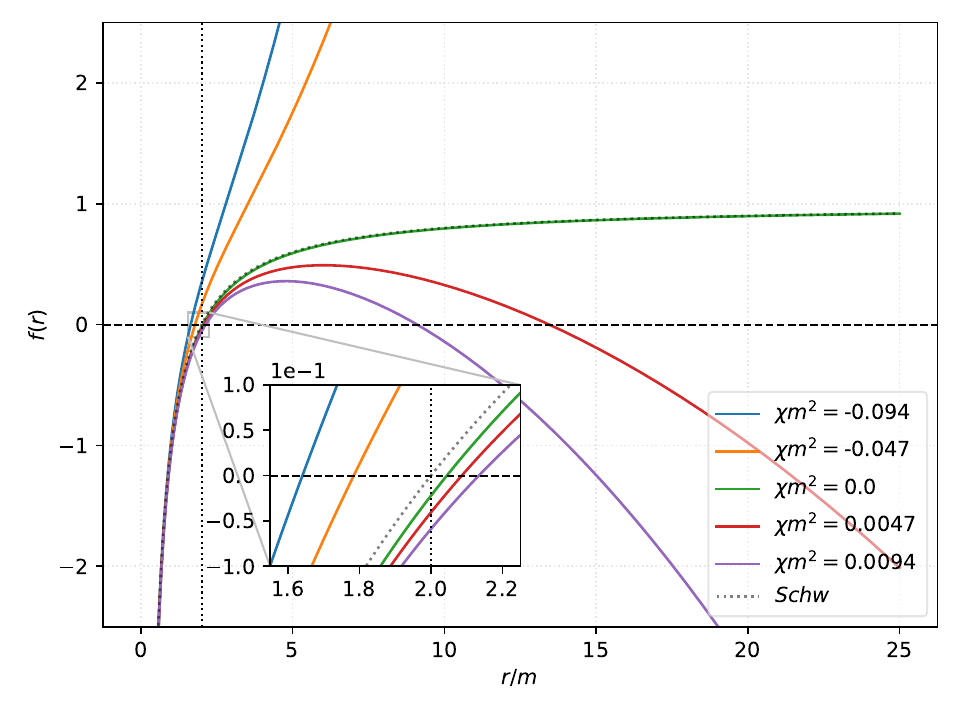}
    \includegraphics[width=0.48\textwidth]{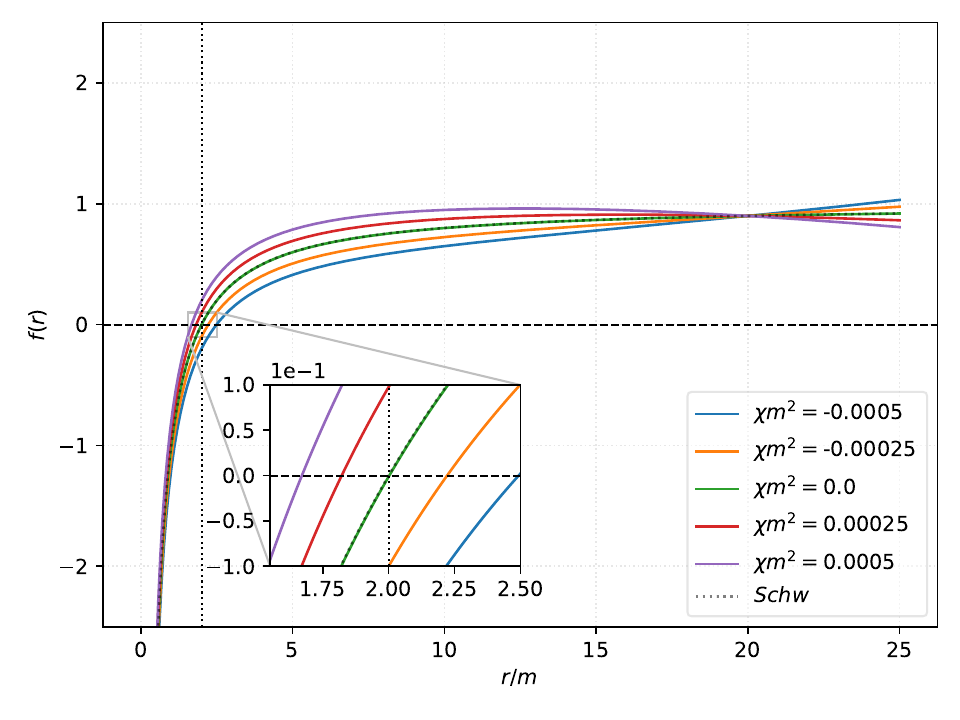}
    \includegraphics[width=0.48\textwidth]{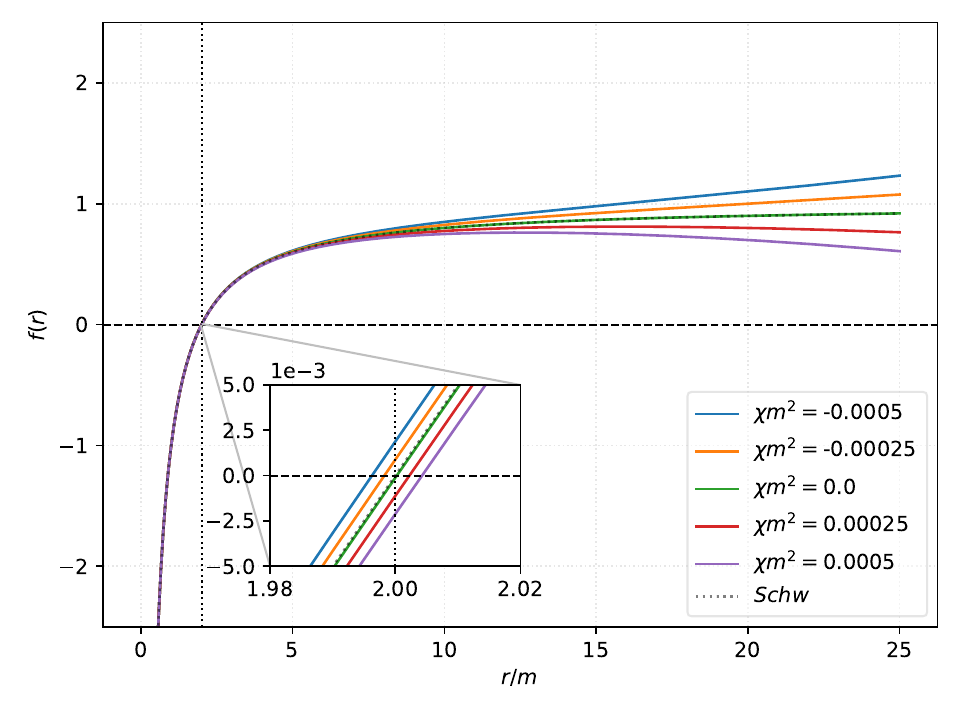}
    \includegraphics[width=0.48\textwidth]{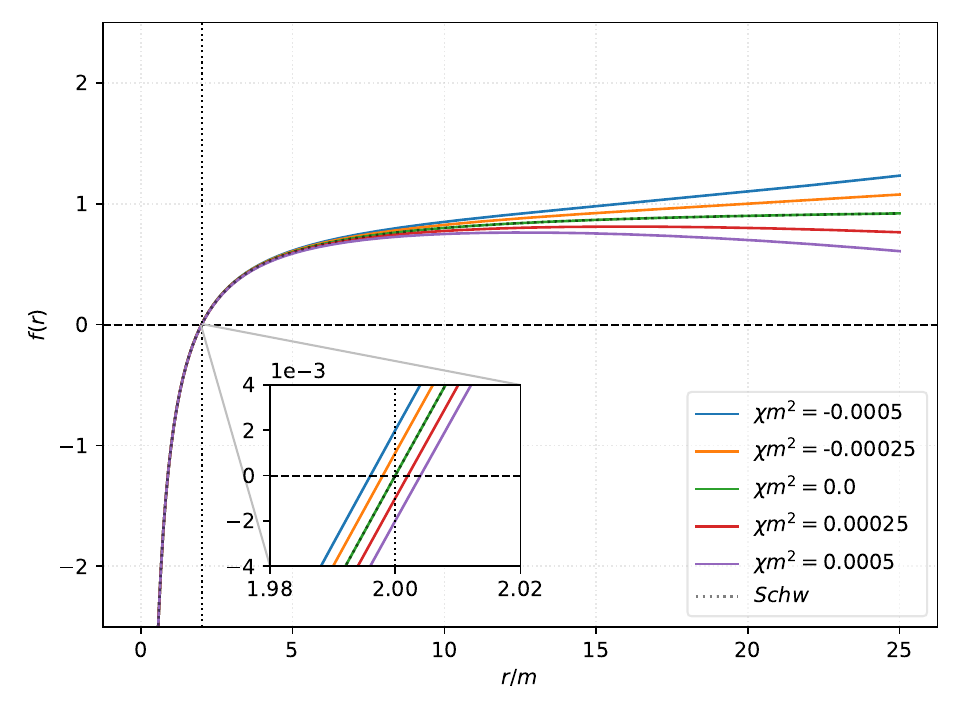}
    \caption{Plot of $F(r) = 0$} to locate the event horizon. The upper left panel corresponds to the black hole with the exponential density profile. The upper right, lower left, and lower right panels correspond to the CDM, SFDM, and URC profiles, respectively. The dotted vertical line corresponds to $r = 2m$ - the horizon for the Schwarzschild case. In this plot, we set some arbitrary values $k/m = 0.01$, and $r_d/m = 20$.
    \label{horizon}
\end{figure*}

In Fig. \ref{horizon}, we plot the lapse/metric function $A(r)$ to locate horizon formation given different values of the GUP parameter $\chi$.
With the same values of the parameters, we can see how the horizon deviates from the standard Schwarzschild case, with different sensitivities to the dark matter profile and the GUP parameter. The green curve represents the case where $\chi = 0$. In this case, $k$ increases the horizon radius for the exponential profile, while the other three profiles show negligible deviation. The deviation becomes large if $k$ is comparable to $m$ or larger. With $\chi < 0$, the horizon shifts to a lower radius except for the CDM profile. Both the exponential and the CDM profiles are sensitive to the effect of the GUP parameter, while the SFDM and URC profiles have shown minimal deviations, as shown in the inset plots. More importantly, the horizon deviates on both signs of $\chi$ \cite{Bosso:2022xnm}, and it also clearly mimics the effect of the cosmological constant since a second horizon is formed.

\section{Null geodesic and the black hole shadow} \label{sec3}
\subsection{Photonsphere}
Since the black hole solutions obtained are static and spherically symmetric, we can study the deviations caused by the different DM profiles and the GUP parameter on the photonsphere at $\theta = \pi/2$ without losing generality. To achieve this, we use a simple but well-known method developed in Ref. \cite{Perlick:2015vta,Perlick:2021aok}.

Consider the Lagrangian
\begin{equation}
    \mathcal{L} = \frac{1}{2}\left( -A(r) \dot{t}^2 +B(r) \dot{r}^2 + r^2 \dot{\phi}^2 \right),
\end{equation}
where a simple implementation of the variational principle leads to the two constants of motion:
\begin{equation}
    E = A(r)\frac{dt}{d\lambda}, \qquad L = r^2\frac{d\phi}{d\lambda}.
\end{equation}
From here, we can define the impact parameter as
\begin{equation}
    b \equiv \frac{L}{E} = \frac{r^2}{A(r)}\frac{d\phi}{dt}.
\end{equation}
For null geodesics, $g_{\mu \nu}\dot{x}^\mu \dot{x}^\nu = 0$, and we obtain the orbit equation as
\begin{equation}
    \left(\frac{dr}{d\phi}\right)^2 =\frac{r^2}{B(r)}\left(\frac{h(r)^2}{b^2}-1\right),
\end{equation}
where the function $h(r)$ is defined as
\begin{equation}
     h(r)^2 = \frac{r^2}{A(r)}.
\end{equation}
The location of the photonsphere $r_\text{ps}$ can be sought off through the condition $h'(r) = 0$, where the prime denotes differentiation with respect to $r$. Albeit no closed-form solution for $r_\text{ps}$ was obtained, we present the results for the exponential profile, CDM, SFDM, and URC, respectively:
\begin{equation}
    \frac{d^{2}}{8}\left(3 m -r \right) - 3\pi  k \left[\left(d^{2}+\frac{2}{3} r d +\frac{1}{6} r^{2}\right) {\mathrm e}^{-\frac{r}{d}}-d^{2}\right]  +\pi  \,r^{2} \chi k\left[\left(d^{2}+r d +\frac{1}{2} r^{2}\right) {\mathrm e}^{-\frac{r}{d}}-d^{2}\right]  = 0,
\end{equation}
\begin{align}
    \left(m -\frac{r}{3}\right) \left(r_d +r \right)&+\frac{k}{3}\left[12 \pi \left(r_d +r \right)  \ln \! \left(1+\frac{r}{r_d}\right)-4 \pi  r \right] -\frac{r r_d^{2} \chi}{3} \left(r_d +r \right) \\ \nonumber
    &+ \frac{ \chi k}{3}\left[4 \left(r_d +r \right) \pi  \left(3 r_d^{2}-r^{2}\right) \ln \! \left(1+\frac{r}{r_d}\right)-4 r \left(r_d +r \right) \left(r_d \pi -\pi  r \right)\right]  = 0,
\end{align}
\begin{equation}
    \pi^{2} \left(m -\frac{r}{3}\right) r_d +\frac{k}{3}\left[12 \sin \! \left(\frac{r \pi}{r_d}\right) r_d - 4 \pi  r \cos \! \left(\frac{r \pi}{r_d}\right) \right]  + \frac{ 4 r^{2} \chi k}{3}\left[\pi  r \cos \! \left(\frac{r \pi}{r_d}\right)-\sin \! \left(\frac{r \pi}{r_d}\right) r_d \right]  = 0,
\end{equation}
\begin{equation}
    \left(m -\frac{r}{3}\right) \left(r_d +r \right) \left(r_d^{2}+r^{2}\right) r_d +\frac{ 2 \pi  rk }{9} \Bigg\{\left(r_d +r \right) \left(r_d^{2}+r^{2}\right) \left[\ln \! \frac{(r+r_d)^2}{r_d^{2}+r^{2}}-2\arctan \! \left(\frac{r}{r_d}\right)\right] + 2r_d \,r^{2}\Bigg\}-\frac{4 \pi  k \,r^{5} r_d \chi}{9} = 0.
\end{equation}

\begin{figure*}
    \centering
    \includegraphics[width=0.48\textwidth]{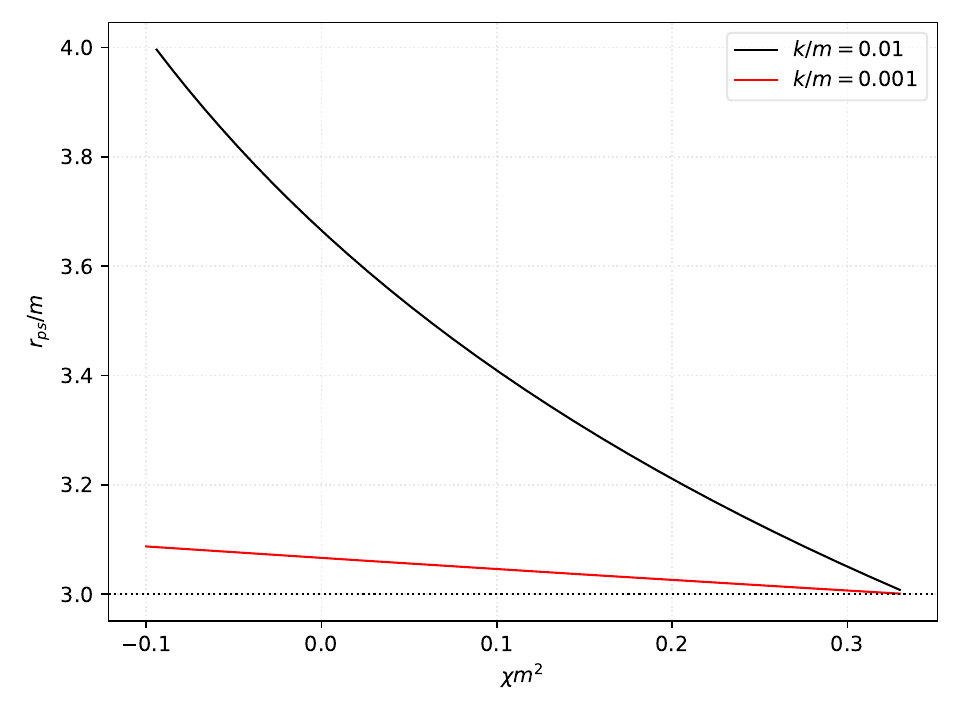}
    \includegraphics[width=0.48\textwidth]{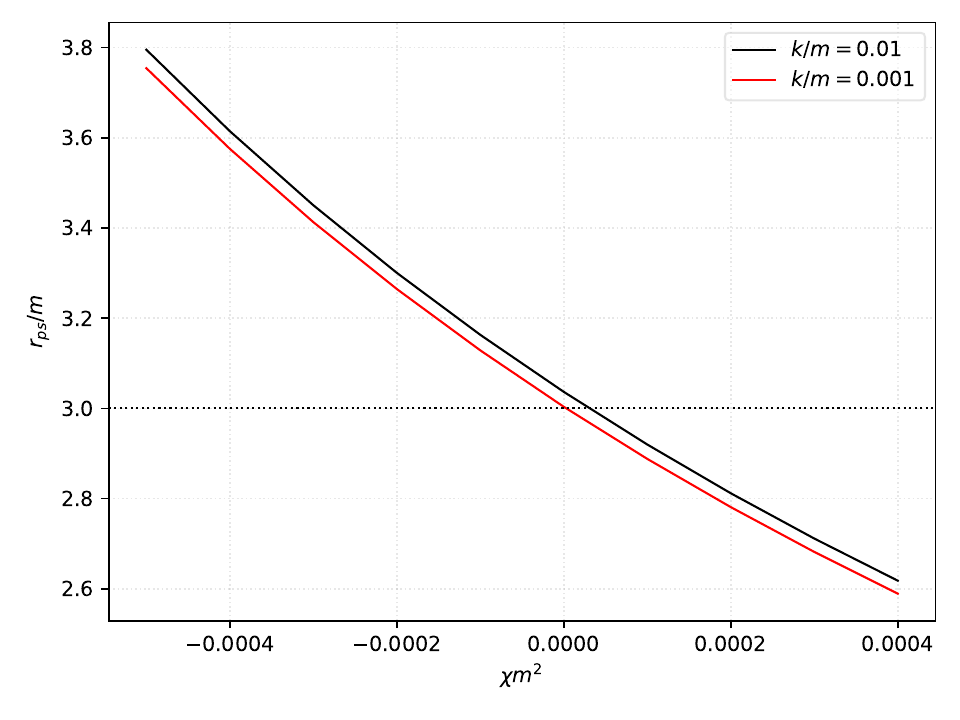}
    \includegraphics[width=0.48\textwidth]{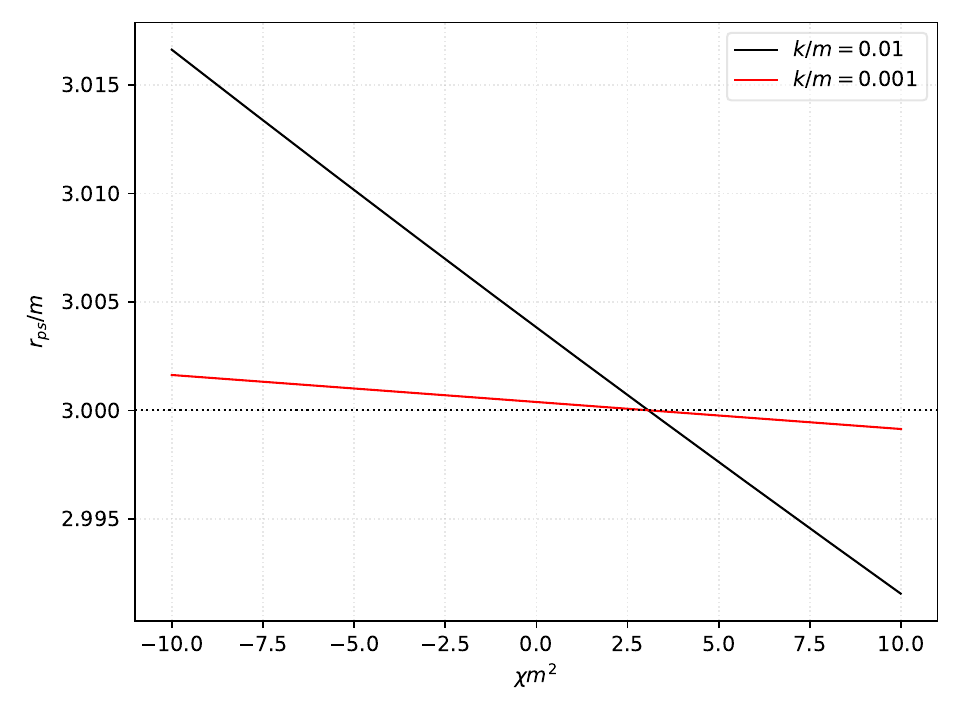}
    \includegraphics[width=0.48\textwidth]{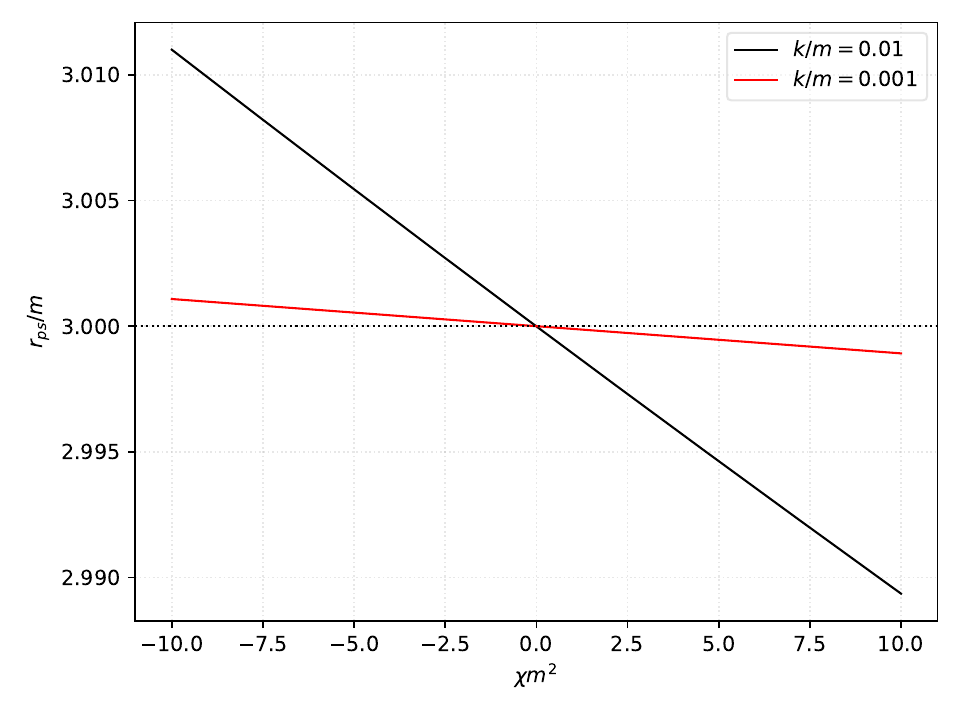}
    \caption{Plot of the photonsphere behavior as $\chi m^2$ changes. The upper left, upper right, lower left, and lower right panels show the black hole with exponential, CDM, SFDM, and URC profiles, respectively. The horizontal dotted lines represent the photonsphere radius in the Schwarzschild case, $r_\text{ps} = 3m$. In this plot, we set some arbitrary values $k/m$, and $r_d/m = 20$.}
    \label{rph}
\end{figure*}

It is possible to study the photonsphere behavior through numerical analysis. The plot is shown in Fig. \ref{rph} for all the dark matter profiles used in this study.
We can see how the photonsphere deviates from the Schwarzschild case due to the different values and signs of $\chi$ and the pure dark matter effect $(\chi = 0)$. For the latter case, the photonsphere tends to increase with $k/m$. Such an effect is strong in the exponential and the SFDM profiles. For the CDM profile, the increase in the photonsphere radius becomes evident as $k/m$ becomes more concentrated relative to $r_d/m$. The effect seems weak to the URC profile. For the former case, we generally see that the photonsphere decreases as $\chi/m$ increases. The radius tends to $r_\text{ps}/m = 3$ in the exponential profile, while the other three profiles permit less than the Schwarzschild case. We also notice the same trend for the SFDM and URC profiles, where the latter profile gives smaller values of $r_\text{ps}/m$.

\subsection{Black hole shadow and constraints}
Having obtained the expression for $h'(r) = 0$ gives us the ability to backward trace the path of photons, which escaped from the photonsphere. For some observer at $r_{o}$, one can define \cite{Perlick:2015vta,Perlick:2021aok}
\begin{equation}
    \tan(\alpha_{\text{sh}}) = \lim_{\Delta x \to 0}\frac{\Delta y}{\Delta x} = \left(\frac{r^2}{B(r)}\right)^{1/2} \frac{d\phi}{dr} \bigg|_{r=r_\text{o}},
\end{equation}
which can be simplified as
\begin{equation}
    \sin^{2}(\alpha_\text{sh}) = \frac{b_\text{crit}^{2}}{h(r_\text{o})^{2}}.
\end{equation}
Here, the critical impact parameter is a function of the photonsphere radius $r_\text{ps}$. It is well-known that if a black hole is not affected by any astrophysical environment, such as the Schwarzschild case, the shadow radius $R_\text{sh}$ is equal to $b_\text{crit}$ However, for black hole spacetimes that are not asymptotically flat, the photons traveling from the photonsphere to the observer at $r_\text{o}$ deviates from the Schwarzschild case due to the astrophysical environment \cite{Abdujabbarov:2015pqp}. The critical impact parameter is given by \cite{Pantig:2022ely}
\begin{equation}
    b_\text{crit}^2 = \frac{h(r)}{B'(r)r^2-2B(r)r} \left(h(r)B'(r)r^2-2h(r)B(r)r-2h'(r)B(r)r^2 \right) \bigg|_{r = r_\text{ps}}.
\end{equation}
Since the resulting expression is rather complicated, we state for the four density profiles that the shadow radius is obtained through
\begin{equation}
    R_\text{sh} = b_\text{crit}\sqrt{A(r_\text{o})}.
\end{equation}

\begin{figure*}
    \centering
    \includegraphics[width=0.48\textwidth]{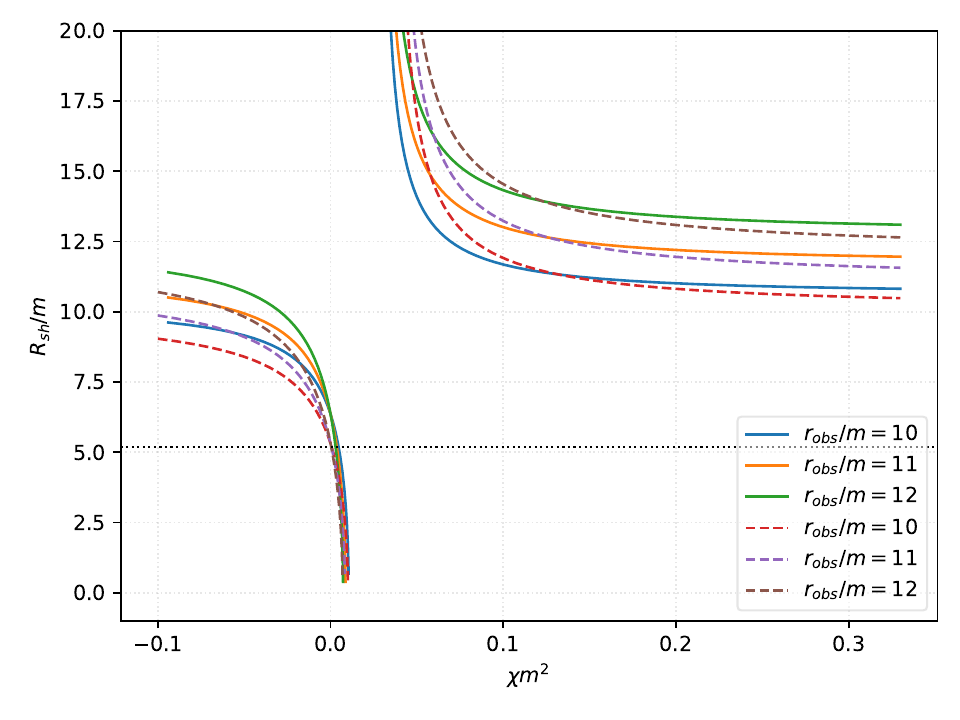}
    \includegraphics[width=0.48\textwidth]{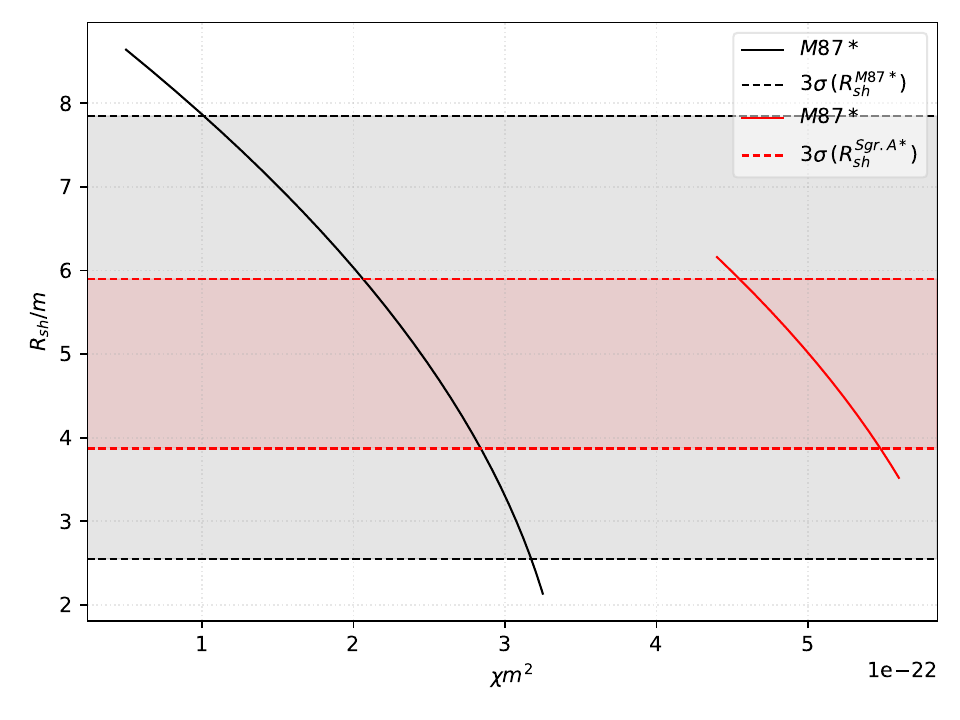}
    \caption{Shadow behavior for the 1st density profile. In the left panel, we show the general trend of the shadow radius, where $k = 0.01$ (solid lines), and $k = 0.001$ (dashed lines). Both have $r_d/m = 20$ as the core radius. The horizontal dotted lines represent the Schwarzschild shadow radius. In the right panel, the upper and lower bounds for $\chi m^2$ are shown for both M87* and Sgr. A* at $3\sigma$ confidence level. For M87*, $k/m = 0.0363, r_d/m = 1.577\times10^{7}$. For Sgr. A*, $k/m = 0.0549$ with $r_d/m = 2.384\times10^{9}$.}
    \label{rsh}
\end{figure*}

\begin{figure*}
    \centering
    \includegraphics[width=0.48\textwidth]{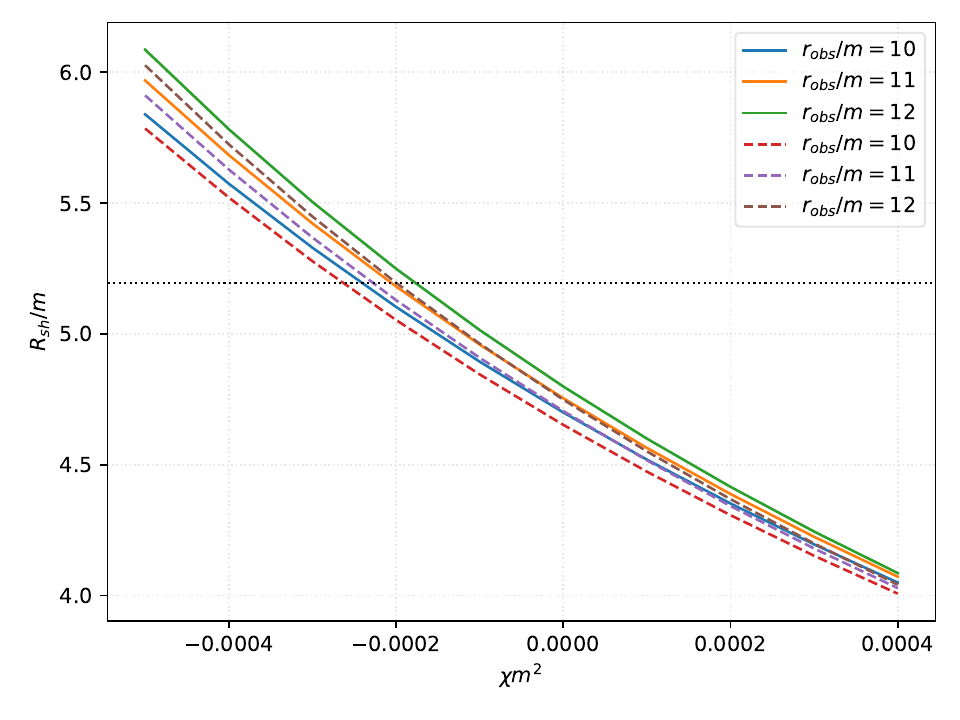}
    \includegraphics[width=0.48\textwidth]{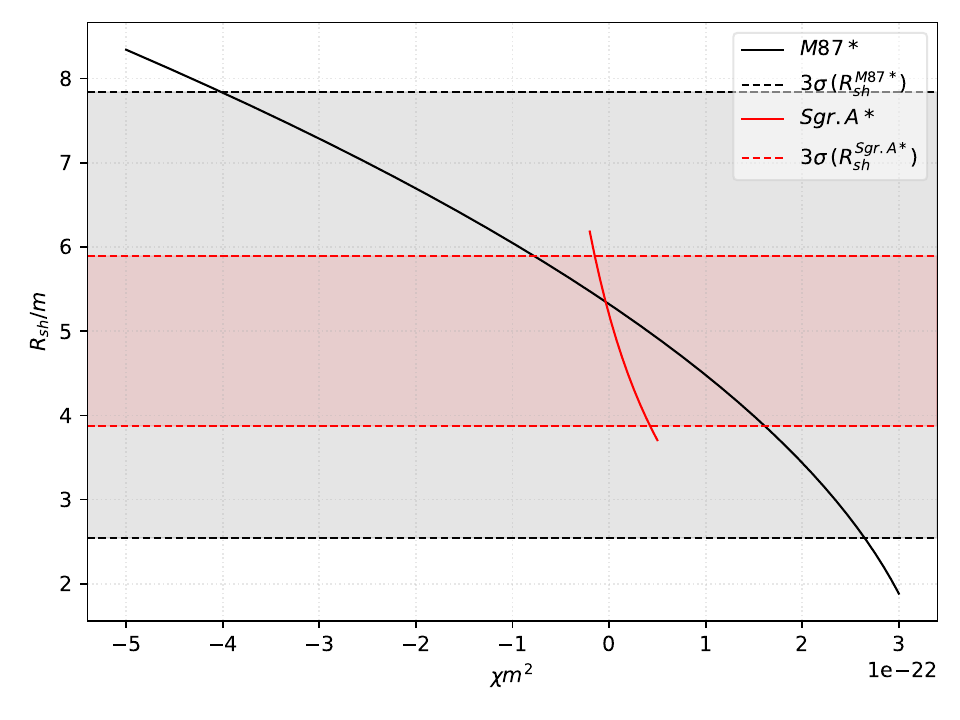}
    \caption{Shadow behavior for the CDM density profile. In the left panel, we show the general trend of the shadow radius, where $k = 0.01$ (solid lines), and $k = 0.001$ (dashed lines). Both have $r_d/m = 20$ as the core radius. The horizontal dotted lines represent the Schwarzschild shadow radius. In the right panel, the upper and lower bounds for $\chi m^2$ are shown for both M87* and Sgr. A* at $3\sigma$ confidence level. For M87*, $k/m = 80.51, r_d/m = 1.281\times10^{5}$. For Sgr. A*, $k/m = 23952$ with $r_d/m = 8.484\times10^{10}$.}
    \label{rsh_cdm}
\end{figure*}

\begin{figure*}
    \centering
    \includegraphics[width=0.48\textwidth]{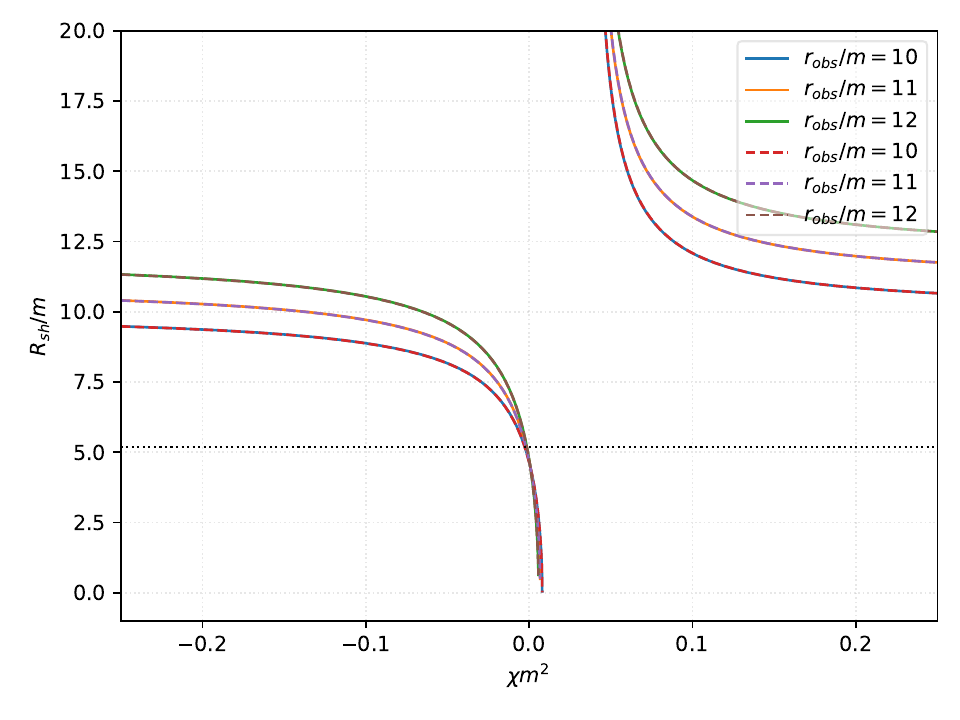}
    \includegraphics[width=0.48\textwidth]{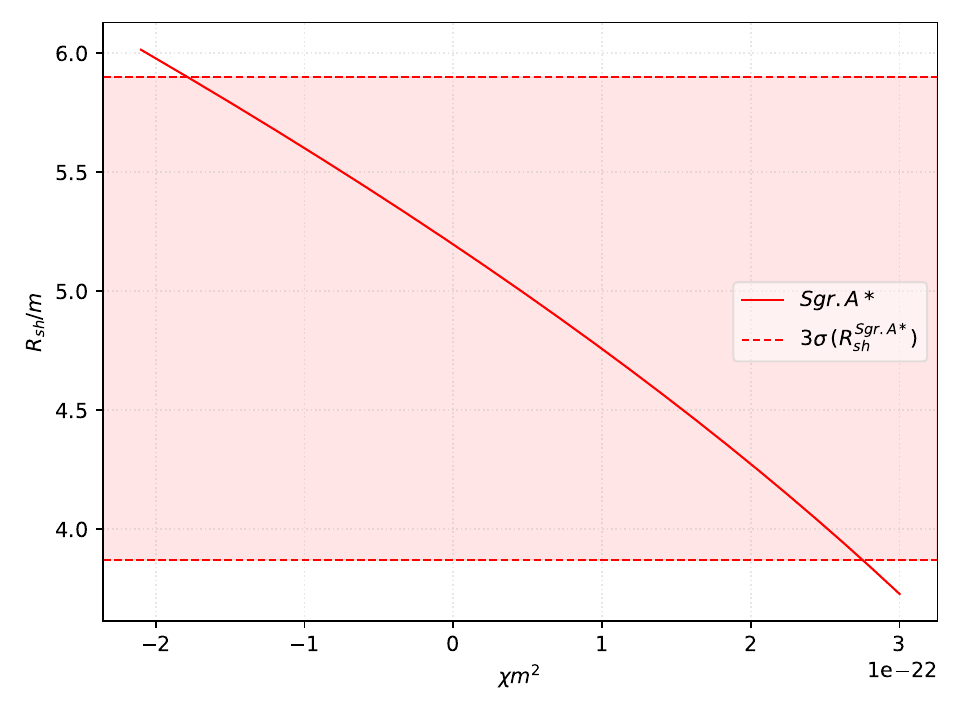}
    \caption{Shadow behavior for the SFDM density profile. In the left panel, we show the general trend of the shadow radius, where $k = 0.01$ (solid lines), and $k = 0.001$ (dashed lines). Both have $r_d/m = 20$ as the core radius. The horizontal dotted lines represent the Schwarzschild shadow radius. In the right panel, the upper and lower bounds $\chi m^2$ are shown Sgr. A* at $3\sigma$ confidence level. Here, $k/m = 30869$ with the same value for $r_d/m = 7.629\times10^{10}$.}
    \label{rsh_sfdm}
\end{figure*}

\begin{figure*}
    \centering
    \includegraphics[width=0.48\textwidth]{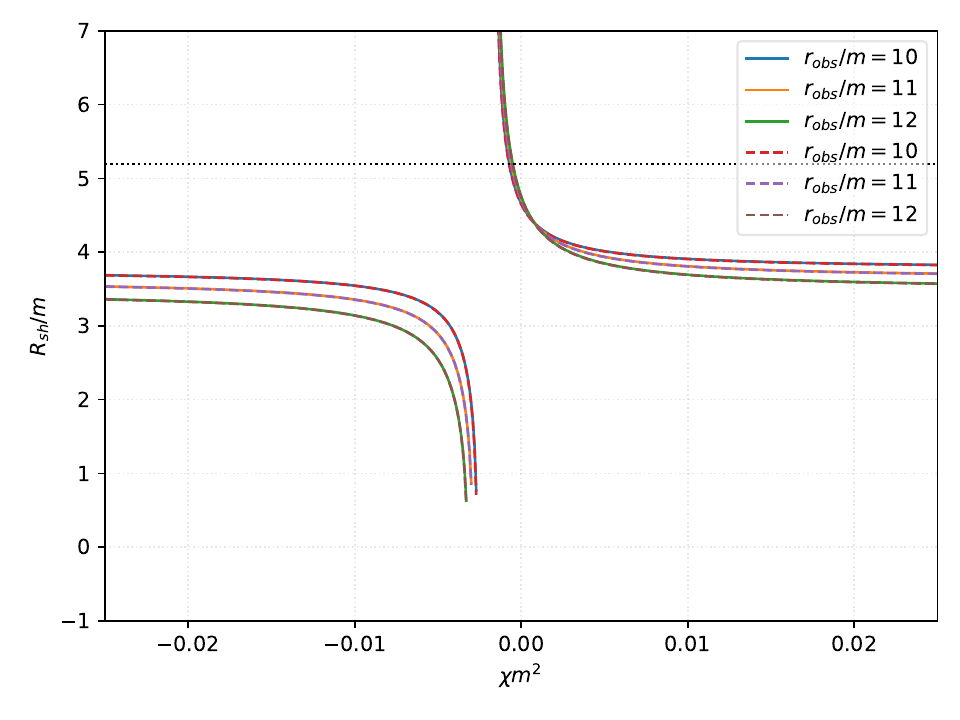}
    \includegraphics[width=0.48\textwidth]{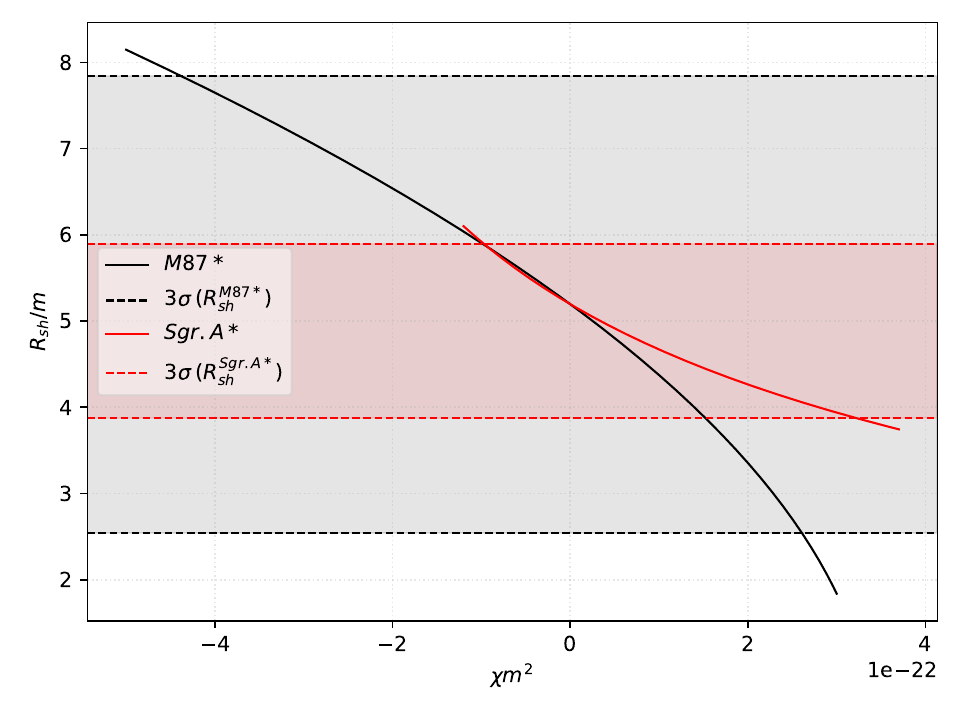}
    \caption{Shadow behavior for the URC density profile. In the left panel, we show the general trend of the shadow radius, where $k = 0.01$ (solid lines), and $k = 0.001$ (dashed lines). Both have $r_d/m = 20$ as the core radius. In the right panel, the upper and lower bounds for $\chi m^2$ are shown for both M87* and Sgr. A* at $3\sigma$ confidence level. For M87*, $k/m = 805, r_d/m = 2.93\times10^{8}$. For Sgr. A*, $k/m = 0.9$ with the same value for $r_d/m = 3.79\times10^{10}$.}
    \label{rsh_urc}
\end{figure*}

The result of the numerical analysis is shown on the left panel of Figs. \ref{rsh}-\ref{rsh_urc}, where we theoretically studied the various trends of the shadow radius as $\chi$ changes for different observer locations and using the values of $r_\text{ps}/m$ obtained to generate Fig. \ref{rph}. Indeed, the behaviors are unique for each of the density profiles. The effect of increasing the observer distance from the black hole makes the shadow radius increase except for the URC profile.  Then, except for the CDM profile, we see asymptotic behaviors for the exponential, SFDM, and URC profiles. The vertical asymptote shows that the shadow radius becomes abnormally large (or small) relative to the Schwarzschild case, which occurs when $\chi$ is close to zero. When $\chi$ becomes more negative or positive, we observe horizontal asymptotic behavior for some value of $R_\text{sh}$. For the CDM profile, we notice that very small values of $\chi$ are needed, and the shadow radius increases as the GUP parameter becomes more negative.

Next, we aim to find constraints to $\chi$ using the EHT data in Table \ref{tab1}.
\begin{table}
    \centering
    \begin{tabular}{ p{2cm} p{3.5cm} p{4.5cm} p{2cm}}
    \hline
    \hline
    Black hole & Mass $m$ ($M_\odot$) & Angular diameter: $2\alpha_\text{sh}$ ($\mu$as) & Distance (kpc) \\
    \hline
    Sgr. A*   & $4.3 \pm 0.013$x$10^6$ (VLTI)    & $48.7 \pm 7$ (EHT) &   $8.277 \pm 0.033$ \\
    M87* &   $6.5 \pm 0.90$x$10^9$  & $42 \pm 3$   & $16800$ \\
    \hline
    \end{tabular}
    \caption{Black hole observational constraints.}
    \label{tab1}
\end{table}
To validate the models used in this study, we apply a simple parameter estimation through a fitting procedure. We want to see the bounds for $\chi$, if they exist, at $3\sigma \,(99.7\%)$ confidence level. Looking at Refs. \cite{EventHorizonTelescope:2019dse,EventHorizonTelescope:2022xnr,EventHorizonTelescope:2021dqv,Vagnozzi:2022moj}, the $3\sigma$ level of significance should read $3.871M \leq R_\text{sh} \leq 5.898M$, and $ 2.546M \leq R_\text{sh} \leq 7.846M$ for Sgr. A* and M87*, respectively.

It is correct that these black holes have spin parameter $a$, and constraining $\chi$ using the non-rotating solution may give erroneous results. However, we will hold to the exposition on Ref. \cite{Vagnozzi:2022moj} about why considering the non-rotating case is enough to find constraints for parameters affecting the black hole geometry. Furthermore, we also ignore a far more realistic case of considering the photons that escaped the gravitational grip from the event horizon. Hence, we consider the silhouette of the invisible black hole shadow (see Ref. \cite{Dokuchaev:2019jqq}), which was commonly studied in several papers in the literature. Considering these caveats, our results are presented on the right panel of Figs. \ref{rsh}-\ref{rsh_urc}.
In obtaining these plots, we theoretically assume that the exponential profile in Eq. \eqref{e1} applies to M87* and Sgr. A* in finding these constraints since there are no available data for $k$ and $r_d$ for this kind of density profile. Afterward, we used the realistic data of $k$ and $r_d$ for the CDM, SFDM, and URC profiles as they are readily available in Refs. \cite{Hou:2018bar,Pantig:2022toh}. The obtained values for $\chi$ were extremely small, regardless of its sign. We listed the constrained values for $\chi$ in Table \ref{tab2}. The allowed bounds differ for each dark matter profile with GUP minimal length scale effects, and we can say that these models fit the data with a simple parameter estimation procedure.

Note that $\chi$ is a function of $\gamma$, which is the GUP parameter itself determined in \cite{Bosso:2022xnm}. In Table \ref{tab2}, we present these values of $\gamma$. In finding these values, let us use the exponential profile. The upper bound for $\chi$ on M87* is $1.003\times10^{-22}$ and with Eq. \eqref{chi}, given that $\omega \sim 7.6850\times10^{-20}$ and $r_*/m \sim 2.15\times10^{9}$, we obtained $\gamma \sim 7.85\times10^{34}$. As for the lower bound, $\gamma \sim 2.47\times10^{34}$. We noticed that the EHT constraints permit both signs for $\gamma$ as indicated by the bounds at $3\sigma$ confidence level for the CDM, SFDM, and URC.
\begin{table}
    \centering
    \begin{tabular}{ccccc}
\hline
\hline
{M87* ($3\sigma$)} &  Exponential profile  &  CDM  &  SFDM  &  URC  \\
\hline
Upper bound  & $7.85\times10^{34}$ & $-3.17\times10^{35}$ & - & $-3.45\times10^{35}$ \\
Lower bound & $2.47\times10^{35}$ & $2.06\times10^{35}$ & - & $2.02\times10^{35}$ \\
\hline
\end{tabular}
\quad
\centering
\begin{tabular}{ccccc}
\hline
\hline
{Sgr. A* ($3\sigma$)} &  Exponential profile  &  CDM  &  SFDM  &  URC  \\
\hline
Upper bound  & $3.55\times10^{35}$ & $-1.17\times10^{34}$ & $-1.41\times10^{35}$ & $-7.76\times10^{34}$ \\
Lower bound & $4.28\times10^{35}$ & $3.21\times10^{34}$ & $2.16\times10^{35}$ & $2.52\times10^{35}$ \\
\hline
\end{tabular}
    \caption{Constrained values for the GUP parameter $\gamma$ in terms of the black hole mass $m$ according to the EHT data of M87* and Sgr. A*.}
    \label{tab2}
\end{table}

\section{Conclusion} \label{sec4}
In this work, we obtained four black hole solutions surrounded with dark matter, which is perceived as a minimal length scale effect. In other words, dark matter is under the effect of GUP, fusing its properties to the black hole geometry. The union allowed the GUP effect to manifest on the black hole shadow, which is the focus of this study. The resulting black hole metrics, albeit spherically symmetric, possess non-asymptotic flatness, which is not seen in other related studies. Indeed, the minimal length scale effect adds to dark matter effects at vast distances, far from the galactic centers where the supermassive black hole resides.

First, the dark matter effect from different models has shown differences in deviation from the Schwarzschild event horizon (see Fig. \ref{horizon}). Then, we theoretically examine how the photonsphere behaves under different dark matter models (see Fig. \ref{rph}). While the SFDM and URC have similar trends, these two differ with the exponential and the CDM profiles. Nonetheless, the four have the same description that $r_\text{ps}$ decreases as $\chi$ becomes more negative. As these photons travel through the astrophysical environment containing dark matter, deviations to the perceived shadow radius occur (see the left panel of Figs. \ref{rsh}-\ref{rsh_urc}). We should note that in these plots, we have chosen arbitrary values of $k/m$, $r_d/m$, and $r_\text{o}/m$ to quickly gain some overview of how the photonsphere and shadow radius behave. 

Next, we used the available EHT data for the angular shadow radii at $3\sigma$ confidence level to find constraints for the GUP parameter $\gamma$ (see Table \ref{tab2}). Such a simple fitting procedure allows us to determine which model is favored by data. Using the available empirical data for $k/m$ and $r_d/m$ for CDM, SFDM, and URC profiles, we have seen they these were qualified for such a parameter estimation if combined with the GUP parameter $\chi$. We observed that the upper bound is negative, while the lower bound is positive. Thus, studying the dark matter effect under the GUP minimal length scale on the black hole shadow confirms claims on estimating the positive bounds for the GUP parameter in laboratory experiments \cite{Pikovski:2011zk, Bosso:2016ycv, Kumar:2017cka} and the negative bounds coming from astrophysical/cosmological observations \cite{Das:2021nbq, Ong:2018zqn, Nenmeli:2021orl, Jizba:2023mof, Buoninfante:2019fwr, Jizba:2022icu}. While less realistic for the exponential profile \cite{Bosso:2022xnm} on its applicability to M87* and Sgr. A* black holes, we saw that the model also fits the EHT constraints for the positive upper and lower bounds, in contrast to the constraint found as $ \gamma / r_*^2 \sim -3.6\times 10^{27} \text{ ly}^{-2}$ \cite{Bosso:2022xnm}. Although this is the case, we remark that extending $\chi$ to be more negative will result in a very large shadow radius that is outside the $3\sigma$ level.

The EHT has already provided us with the first direct image of a black hole, but it is still limited in its resolution and sensitivity. More sophisticated technologies could allow us to probe black hole's geometry in more detail, revealing imprints of what certain parameters (in this paper's case, the GUP parameter $\chi$) caused the uncertainties from the mean measurement. For instance, more powerful radio telescopes could be used to image black holes with higher resolution and sensitivity. This would allow us to see finer details of the black hole's structure and to measure its properties more accurately. Gravitational wave observatories like LIGO and Virgo can detect the gravitational waves emitted by black holes when they merge. By studying the gravitational waves from black hole mergers, astronomers can learn more about the physics of black holes and test different theories of black hole physics. Finally, space-based observatories could also study black holes at X-ray and gamma-ray wavelengths. One example, in principle, is the Solar System-based Very Long Baseline Interferometry.

We direct as future research the possibility of extending the study to a more realistic case of rotating black holes. However, this could be a challenge since the Newman-Janis algorithm does not simply work in non-asymptotically flat spacetime, where properties mimic the effect of the cosmological constant. Eq. \eqref{emetric} is general for any distance-dependent expression/model for $\beta$, which can describe the GUP effect on rotation curves. It is interesting to think of such a model and even test it on other dark matter profiles.

\begin{acknowledgements}
R. P. and A. {\"O}. would like to acknowledge networking support by the COST Action CA18108 - Quantum gravity phenomenology in the multi-messenger approach (QG-MM). A. {\"O}. would like to acknowledge networking support by the COST Action CA21106 - COSMIC WISPers in the Dark Universe: Theory, astrophysics and experiments (CosmicWISPers).
\end{acknowledgements}

\bibliography{ref}
\end{document}